%% file: sigir26-ads-in-rag-frame.tex
\documentclass[sigconf, arxiv]{acmart}
\pdfoutput=1 

\input{sigir26-ads-in-rag.sty}

\newcommand{\bswebis}[1][ ]{} 
\newcommand{\bswebisuc}[1][ ]{} 

\begin{document}
\input{sigir26-ads-in-rag-pre}
\input{sigir26-ads-in-rag-part1}

\input{sigir26-ads-in-rag-part2}
\input{sigir26-ads-in-rag-part3}
\input{sigir26-ads-in-rag-part4}
\input{sigir26-ads-in-rag-part5}
\input{sigir26-ads-in-rag-sum}

\bibliographystyle{ACM-Reference-Format}
\bibliography{sigir26-ads-in-rag-lit}
\end{document}

%% file: sigir26-ads-in-rag-pre.tex
\title{Detecting RAG Advertisements Across Advertising Styles}
\settopmatter{authorsperrow=4}

\author[S.\ Heineking]{Sebastian Heineking}
  \affiliation{
  \institution{University of Kassel}
  \city{Kassel}
  \country{Germany}
}
\orcid{0000-0002-7701-3294}

\author[W.\ Pertsch]{Wilhelm Pertsch}
\affiliation{
  \institution{Friedrich-Schiller-Universit{\"a}t Jena}
  \city{Jena}
  \country{Germany}
}
\orcid{0009-0008-9232-4788}

\author[I.\ Zelch]{Ines Zelch}
  \affiliation{
  \institution{Friedrich-Schiller-Universit{\"a}t Jena}
  \city{Jena}
  \country{Germany}
}
\orcid{0009-0005-2659-5326}

\author[J.\ Bevendorff]{Janek Bevendorff}
\affiliation{
  \institution{Bauhaus-Universit{\"a}t Weimar}
  \city{Weimar}
  \country{Germany}
}
\orcid{0000-0002-3797-0559}

\author[B.\ Stein]{Benno Stein}
\affiliation{
  \institution{Bauhaus-Universit{\"a}t Weimar}
  \city{Weimar}
  \country{Germany}
}
\orcid{0000-0001-9033-2217}

\author[M.\ Hagen]{Matthias Hagen}
  \affiliation{
  \institution{Friedrich-Schiller-Universit{\"a}t Jena}
  \city{Jena}
  \country{Germany}
}
\orcid{0000-0002-9733-2890}

\author[M.\ Potthast]{Martin Potthast}
\affiliation{
  \institution{University of Kassel, hessian.AI, and ScaDS.AI}
  \city{Kassel}
  \country{Germany}
}
\orcid{0000-0003-2451-0665}

\renewcommand{\shortauthors}{Sebastian Heineking et al.}

\hyphenation{Modern-BERT}

\begin{abstract}
Large language models (LLMs) enable a new form of advertising for retrieval-augmented generation~(RAG) systems in which organic responses are blended with contextually relevant ads. The prospect of such ``generated native ads'' has sparked interest in whether they can be detected automatically. Existing datasets, however, do not reflect the diversity of advertising styles discussed in the marketing literature. 
In this paper, we
\Ni
develop a taxonomy of advertising styles for LLMs, combining the style dimensions of explicitness and type of appeal,
\Nii
simulate that advertisers may attempt to evade detection by changing their advertising style, and
\Niii 
evaluate a variety of ad-detection approaches with respect to their robustness under these changes. 
Expanding previous work on ad detection, we train models that use entity recognition to exactly locate an ad in an LLM response and find them to be both very effective at detecting responses with ads and largely robust to changes in the advertising style. Since ad blocking will be performed on low-resource end-user devices, we include lightweight models like random forests and SVMs in our evaluation. These models, however, are brittle under such changes, highlighting the need for further efficiency-oriented research for a practical approach to blocking of generated~ads.%
\footnote{Code and data: \url{https://anonymous.4open.science/r/detecting-rag-advertising-styles/}}
\end{abstract}

\ccsdesc[500]{Information systems~Content match advertising}

\keywords{Online Advertising; Retrieval-augmented Generation; Large Language Models}

\copyrightyear{2026}
\acmYear{2026}
\setcopyright{cc}
\setcctype{by}

\makeatletter
\if@ACM@anonymous

\gdef\addresses{\@author{\vskip9ex Anonymous Author(s) \vskip9ex%
        \ifx\@acmSubmissionID\@empty\else\\Submission Id:
          \@acmSubmissionID\fi}}
\renewcommand{\bsacmplace}{January, 2026}
\renewcommand{\bswebis}[1][ ]{}
\renewcommand{\bswebisuc}[1][ ]{}
\acmConference[SIGIR '26]{Proceedings of the 49th International ACM SIGIR Conference on Research and Development in Information Retrieval}{July 20--24, 2026}{Melbourne, Australia}
\acmBooktitle{Proceedings of the 49th International ACM SIGIR Conference on Research and Development in Information Retrieval (SIGIR '26), July 20--24, 2026, Melbourne, Australia}

\else

\renewcommand{\bswebis}[1][ ]{Webis#1}
\renewcommand{\bswebisuc}[1][ ]{WEBIS#1}

\renewcommand{\bsacmplace}{January 2026}
\acmConference[ArXiv]{}{January 2026}{\href{https://webis.de}{webis.de}}
\acmBooktitle{ArXiv}
\acmDOI{} 
\acmISBN{}


\fi
\makeatother

\maketitle

%% file: sigir26-ads-in-rag-part1.tex
\section{Introduction}

Ads will soon be added to responses of large language models. Commercial chatbots such as ChatGPT started as free services, but vendors quickly sold subscriptions to Pro features to offset their vast expenses. Yet only~5\,\% of OpenAI's users are paying subscribers~\cite{sultan:2025} and others are likely not better off. All vendors are therefore working on integrating ads into their chatbots, as evidenced by press releases~\cite{perplexity:2024,sainsbury-carter:2025,openai:2026}, research~\cite{dutting:2024,hajiaghayi:2024}, and industry news~\cite{adegbola:2025,goodwin:2025}. An open question is what \emph{form} these ads will take.

\input{figures-and-tables/figure-2x2-matrix}

In this paper, we study a new form of so-called ``generated native advertising'' (Figure~\ref{fig:2x2-matrix}), in which LLMs blend ads directly into otherwise organically generated responses~\cite{zelch:2024}. Native advertising predates the digital age and has to date been associated primarily with journalistic media such as newspapers and magazines~\cite{amazeen:2020,wojdynski:2016}. Crucially, native ads are deliberately designed by publishers and advertisers to look and read like genuine news articles and to create the impression of editorial content. Even though regulations in many jurisdictions require every form of advertising to be disclosed to consumers~\cite{porlezza:2017}, such disclosures need not be prominent. And since native ads are more effective when readers remain unaware of them~\cite{schauster:2016}, some publishers render disclosures as inconspicuous as possible, for example, through fine print, low-contrast labels, or unusual formatting.  Native advertising has therefore been criticized as a threat to the credibility of editorial content~\cite{schauster:2016}. As RAG~systems and chatbots have become a popular alternative to conventional search engines~\cite{sor:2025}, and as users grow accustomed to and become less scrupulous of their ostensibly organic responses~\cite{leiser:2024}, introducing native ads into them is a worrying prospect.

We study the robustness of detectors for generated native ads in RAG responses. As research on generated ad detection is still in its infancy (Section~\ref{sec:related_work}), we observe several shortcomings in related work in terms of realism and grounding in marketing research. Our contributions are 
\Ni
a taxonomy of advertising styles for LLMs derived from marketing literature, which encompasses the explicitness of the promotion and the type of appeal as distinguishing characteristics.
This enables
\Nii
a ``simulation'' of how advertisers may attempt to evade native ad detection by changing its style, engaging in a cat-and-mouse game with developers of ad detectors. We compile a suite of test sets and use them to
\Niii
evaluate how robust various types of ad detectors are under domain shifts caused by switching the ad-generating LLM and/or the advertising style.

%% file: figures-and-tables/figure-2x2-matrix.tex
\begin{figure}
\Description{
The figure shows a two-by-two matrix of four RAG responses with advertisements. Above the matrix is a user query that asks ``Are there specific destinations that usually have last minute deals?''.
The first dimension of the matrix is the level of overtness, distinguishing between covert advertisements and overt ones. The second dimension is the type of appeal, i.e., does the advertisement make a rational appeal for the advertised product or does it appeal to emotions?
In each cell is a response containing an advertisement generated for based on the style that results from the combination of the two dimensions.  
}
\centering
\begin{minipage}{0.46\textwidth}
\includegraphics[width=\linewidth]{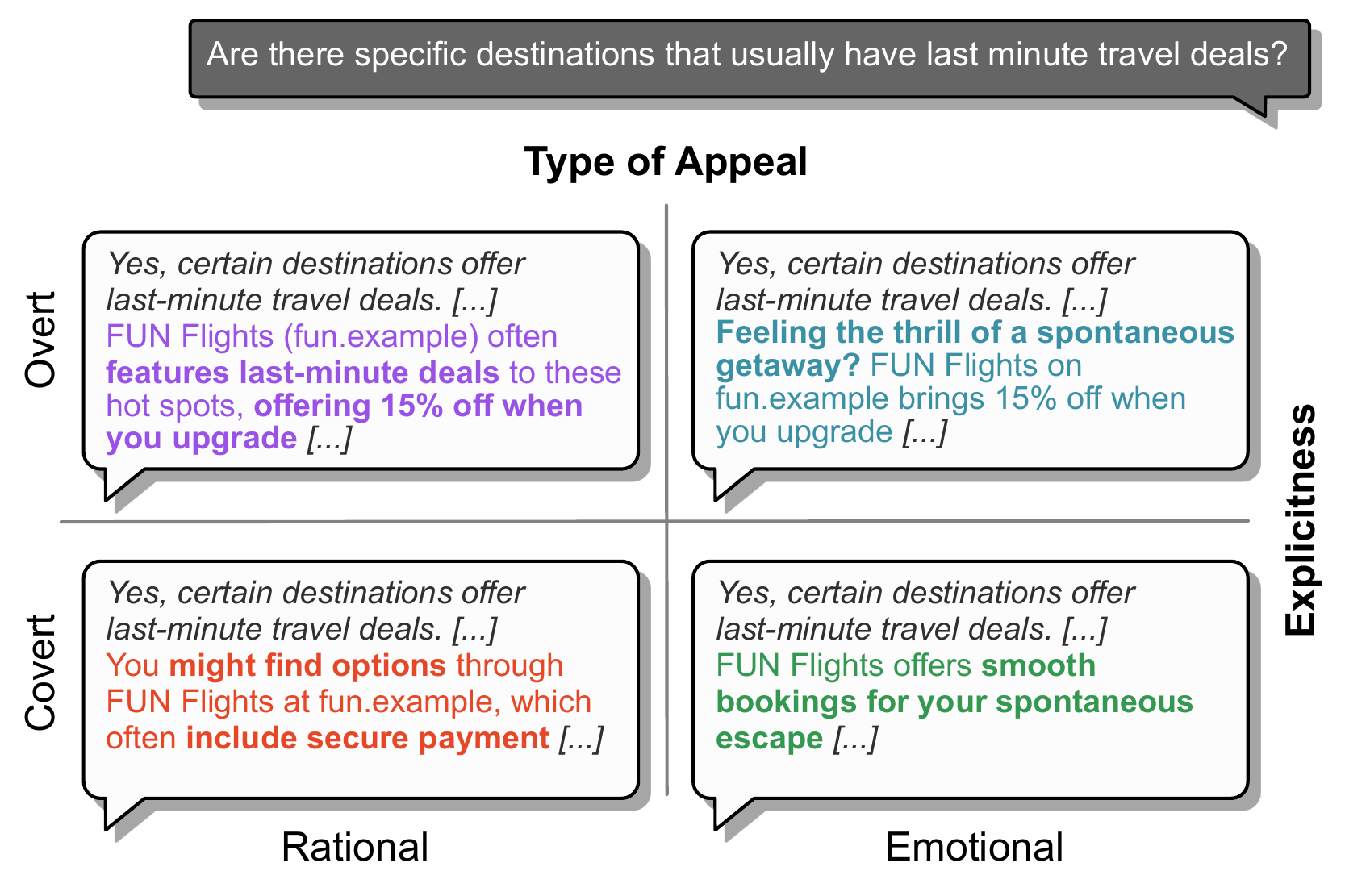}
\end{minipage}
\caption{%
Examples of generated native ads in RAG responses using four advertising styles (one per cell). Note the explicitness of the ad snippets in the ''Overt'' row, or the appeal to emotions in the ``Emotional'' column (marked in bold).
}
\label{fig:2x2-matrix}
\end{figure}

%% file: sigir26-ads-in-rag-part2.tex
\section{Related Work}
\label{sec:related_work}

Generated advertising and its blocking are new fields of research. Below, we summarize the related work on ad auction mechanisms for LLMs, the blending of ads and organic responses, and their detection and blocking.

\subsection{Advertising in LLM Responses}

Advertising in LLM responses has not been deployed so far. Current research focuses primarily on how to design the auction mechanism for interested advertisers and, related to that, the placement of the ads in the responses. The suggested approaches can be divided into ones in which advertisers bid for a specific position in the response~\cite{hajiaghayi:2024,dubey:2024}, and others in which the bids influence the distribution from which tokens are sampled~\cite{dutting:2024,soumalias:2024}. The former treat an ad as a unit placed somewhere in the response. This ensures at least some separation of organic text and advertisements. The latter blur the boundary between organic content and advertising, so that potentially the entire response is sampled from a distribution biased in the interest of advertisers.

As an example of the first scenario, \citet{hajiaghayi:2024} propose an auction model for individual segments, e.g., paragraphs, in a RAG response. Similar to the sponsored links in a classic SERP, advertisements are selected based on query relevance, advertiser bid, and click probability, and added to the context of the LLM that generates the segment.
In the framework proposed by \citet{dubey:2024}, advertisers do not bid on a specific segment but for increased prominence. The authors define prominence as an abstract concept that has a monotonically positive relation to user attention. Higher prominence can mean that an ad is represented with a larger number of words or that it is positioned more visibly in the response. An auction module receives bids, quality scores for the ads, and predicted click-through rates. Through its monotonic allocation function (higher bids result in higher prominence), the auction module outputs a ``prominence allocation'' and prices. The LLM is then instructed to create a text with appropriately prominent ads.

\citet{dutting:2024} present a possible implementation of the second scenario: Advertisers can bid to influence the probability distribution used to sample tokens. Central to their approach are LLM agents that are able to generate advertisements adapted to a specific user query for a specific advertiser. But instead of query-level auctions in which agents generate full responses to queries, the authors propose a token auction model. Within this model, multiple agents jointly generate a response by bidding on their desired probability distributions for the next token, which are aggregated to generate the final response. The higher an agent's bid, the more strongly it affects the aggregated token distribution.
\citet{soumalias:2024} suggest a similar approach. Their framework consists of a reference LLM that generates responses to maximize user satisfaction. Each advertiser is represented by a reward function that takes in a tuple of user query and generated response, and outputs a reward value for the given advertiser. The auctioneer optimizes the final token distribution to produce a response that maximizes the aggregated rewards across all advertisers. This optimization is constrained by a hyperparameter defining the maximum allowed deviation from the original distribution of the reference LLM. This is to ensure the interest of the user is considered as well.

\subsection{Detection of LLM Advertisements}

As discussed in the previous section, LLMs can take predefined messaging, like a brand slogan or claims about a product, and tailor it to a specific user interaction~\cite{dutting:2024,schmidt:2024}. This creates native advertisements that are hard to distinguish from the rest of the ``organic'' response text~\cite{zelch:2024, amazeen:2020}. To the best of our knowledge, there currently exist no publicly available datasets of real LLM-generated advertisements, but this should not hinder research on the topic. \citet{schmidt:2024} created a dataset of synthetic examples to study the detection of potential ads in the responses of conversational search engines. The Webis Generated Native Ads~2024 dataset contains 11{,}303 responses collected from Microsoft Copilot and YouChat. The authors used GPT-4 to insert advertisements into 6{,}401 of these responses. For the identification of the ads, they tested two fine-tuned sentence transformer models, MiniLM and MPNET, and several LLMs for zero-shot classification.

A new version of the dataset, Webis Generated Native Ads~2025 (discussed in Section~\ref{sec:experiment}), was constructed for a shared task, in which participants were asked for approaches to generate, but also to detect ads in LLM responses~\cite{kiesel:2025b}.
For the detection task, the participating groups submitted various transformer models like MPNet and DeBERTa, a random-forest classifier, and logistic regression with TF-IDF features. Across all approaches, the transformer models were the most effective.

A different yet related form of advertising for which organic and publicly available datasets exist, are sponsored segments in videos and podcasts~\cite{reddy:2021,kok-shun:2025}. Similar to how LLMs can be prompted to produce advertising, content creators receive outlines from advertisers, adapt them to their style and audience, and integrate the advertisements into their content. Related work on detecting these types of ads was published by \citet{reddy:2021}, who used transcripts and descriptions of podcasts to detect this kind of ``extraneous content.'' The texts were split into sentences and were classified with BERT and non-neural classifiers using TF-IDF unigram and bigram features.
\citet{kok-shun:2025} prompted GPT-4o on transcripts of YouTube videos to detect segments with sponsored content to assign them the labels ``Media'' or ``Product.'' \citet{bevendorff:2024c} also used transcripts to distinguish between real product reviews and commercial spam content using an SVM and POS~features.

%% file: sigir26-ads-in-rag-part3.tex
\section{The Style of Advertising Language}
\label{sec:experiment}

Marketing research on advertising distinguishes several styles of ads. However, research that pertains particularly to advertising language is relatively scarce, since long-form texts are not the main medium of marketing. As a basis for the emerging research on advertising in LLMs, we derive an operationalizable taxonomy of styles of advertising language. We first analyze the relevant related work from marketing and devise a basic model of how advertisers may go about creating such ads using large language models.


\subsection{Dichotomies of Advertising Styles}
\label{sec:advertising_styles}

In marketing, different schools of thought exist for categorizing advertising styles, all of which are dichotomies, separating advertising into two broad classes. Table~\ref{table-advertising-dichotomies} gives a conceptual overview.
One of them distinguishes between \emph{hard-sell} and \emph{soft-sell} advertising~\cite{okazaki:2010, mueller:1987}. Hard-sell advertising relies on direct promotion, explicitly highlighting and reasoning about positive product attributes. Soft-sell advertising, in contrast, uses indirect communication, invoking images or a specific atmosphere with the goal of appealing to the recipient's emotions. This binary differentiation mixes a two dimensions that are separate in other works with more granular categories: The \emph{explicitness} (or directness) of a promotion and the \emph{type of appeal} it uses.

\input{figures-and-tables/table-advertising-dichotomies}

Ad appeals have been separated into \emph{rational} and \emph{emotional}~\cite{kotler:2018}, or \emph{informational} and \emph{transformational}~\cite{nuweihed:2024, cadet:2017}. A rational appeal highlights positive attributes of a product (e.g., quality, value, performance) to address the audience's self-interest, while an emotional appeal targets their feelings~\cite{kotler:2018}. Likewise, informational ads highlight the usefulness of the products based on facts and reasoning, while transformational ads appeal to the consumer's senses, imagination, and emotions~\cite{nuweihed:2024, cadet:2017}. A potential third category, \emph{moral} appeal~\cite{kotler:2018}, addresses the audience's beliefs about what is ``right.''

We observe conceptual overlap between these dichotomies. However, the differentiations made are highly discipline-specific. We must also take into account that different marketing sub-disciplines have different desiderata. For the purposes of operationalization, however, we blurr out the details and adopt the terminology of the latest dichotomy of rational versus emotional appeal. Combined with the following one, we cover the most salient aspects of marketing styles that are being distinguished.

Ad explicitness has been divided into \emph{overt} and \emph{covert} advertising~\cite{chan:2019}. Overt advertising applies ``traditional'' methods like billboards or banners on web pages. Covert advertising is implemented more subtly, for example in the form of product placement in news articles, editorial content (advertorials), influencer marketing~\cite{vrontis:2021}, or required ingredients in recipes~\cite{luth:2017,chan:2019}. 
The common denominator is that the promotional intent is concealed, for example, by assuming the same appearance as organic, non-advertising content, or by placing the product into a scene without explicity mentioning its name. This makes covert often difficult to recognize~\cite{pierre:2024, goebel:2017, wojdynski:2020}. Studies show that covert ads often (questionably) lead to more positive customer reception~\cite{chan:2019, davtyan:2017}. Some authors also distinguish explicit and implicit advertising, but this seems to be restricted to the context of environmental sustainability~\cite{gong:2023}.

\subsection{Operationalizing Advertising Styles}

The existing research on generated advertising and ad blocking focuses on technical questions such as auction mechanisms, token distributions, or classifier design. Consequently, advertisements are reduced to a type of text that is either inserted into a response or detected to be blocked.
Advertisements, however, are not homogeneous. Instead, there is the aforementioned discourse in marketing research about how to categorize advertising styles. This discourse might be less relevant to conventional ad blocking that relies on URL filter lists, request patterns, or JavaScript behavior to identify advertisements. Ad blockers for LLMs, in contrast to that, will need to identify advertisements based on textual clues alone. As different advertising styles can produce advertisements with different vocabularies and semantics, an understanding of these styles is important for the blocking of generated ads. Therefore, based on our above analysis, we propose a taxonomy of advertising styles for LLMs that is illustrated with the examples in Figure~\ref{fig:2x2-matrix}.

The first dimension of our taxonomy is the \emph{level of explicitness}, distinguishing between \emph{covert} and \emph{overt} advertising. Applied to generated native advertising, this dimension is related to the concept of prominence introduced by~\citet{dubey:2024}. The more overt a generated advertisement is, the more attention it aims to attract by, for example, using a very positive vocabulary or assigning a large share of the response text to the advertisement. Consequently, covert generated advertisements do not seek to attract attention, but rather to influence the recipient in a more subltle way, for instance, by mentioning a product as one possible option among several alternatives.

The taxonomy's second dimension is the \emph{type of appeal}, that can be either \emph{emotional} or \emph{rational}. Advertisements with a rational appeal list (measurable) features about a product to convince their audience. Examples include lower prices than competitors, longer battery life of electronic devices, or important nutrients in food and beverages.
Advertisements with an emotional appeal are more abstract and try to invoke an emotional reaction in the audience. These emotions can be positive like joy, nostalgia, or pride, but also negative like fear, for example in insurance advertising, or sympathy in the case of charity advertising. 
The rational appeal can directly be applied to generated advertising: Instead of listing features about a product on a billboard, they appear in the response. Generated advertisements with an emotional appeal, however, are restricted to short text as a medium to invoke emotions, while other forms of advertising can do so with music or visuals.


\section{Robustness of Ad Detectors}
\label{subsec:experiment_robustness}

Detectors for LLM advertisements rely on patterns in the response text to identify an ad~\cite{schmidt:2024}. These patterns, however, could change depending on the advertising style or LLM used to generate an advertisement. Especially if advertisers actively work against detection by applying their expert knowledge of advertising styles. This raises the question how robust detectors are to changes in how the advertisements are generated. 

\subsection{Simulating Ad Blocking Evasion}

To answer this question, we trained a set of detectors, i.e., classifiers, on a publicly available dataset of LLM responses with advertisements, as a real-world ad blocking developer would. Then, we created responses with new advertisements using our taxonomy of advertising styles as if advertisers were manually trying to avoid the styles used for training the ad blockers and evaluated the classifiers on new test sets with these responses. To measure robustness, we compared the effectiveness of the classifiers on the test split of the publicly available dataset (the \emph{reference test set}) against the effectiveness on our newly generated test sets.

\paragraph{Training Data with a Naive Advertising Style}
\label{paragraph:experiment_dataset}

\input{figures-and-tables/figure-ad-examples}
\input{figures-and-tables/table-dataset-overview}

For our experiments, we used the Webis Generated Native Ads 2025 (WGNA~25) dataset.%
\footnote{\url{https://doi.org/10.5281/zenodo.17830870}.}
The dataset contains 30,731~responses collected from search engines that use retrieval-augmented generation (RAG). The responses were generated by \emph{Brave Search}, \emph{Microsoft Copilot}, \emph{Perplexity}, and \emph{YouChat} in response to 9,062~queries. The majority of queries have a ``commercial character'', as they are based on keywords that a lot of advertisers compete for. Example queries are ``\emph{What are good last minute travel deals?}'' or ``\emph{How do I choose the right size for boys shorts?}''.

For 13,996~of these responses, the dataset creators prompted different LLMs to insert an advertisement. The input data for the advertisements was collected by sending the 9,062~queries to startpage.com and scraping the sponsored results. This resulted in 11,613~items (products or services) with a set of qualities, i.e. claims about the item, and a corresponding advertiser. An example item is given in Figure~\ref{fig:ad-examples}.
Using five different prompts, the dataset creators instructed \mbox{\texttt{gpt-4o}} and \mbox{\texttt{gpt-4o-mini}}, \mbox{\texttt{deepseek-r1}}, \mbox{\texttt{llama3}} and \mbox{\texttt{llama-3.3}} with 70B~parameters, and \mbox{\texttt{qwen-2.5-32b}} to insert advertisements into the 13,996~responses. In total, the dataset contains 44,727~responses divided across train, validation, and test split as illustrated in Table~\ref{tab:dataset_overview}.

\paragraph{Test Data with Evasive Advertising Styles}

\input{figures-and-tables/figure-prompt-example}

The starting point for our experiments is the test split of the WGNA~25 dataset. In the following, we will refer to it as the \emph{reference test set}. It contains 4,316~responses without and 1,904~responses with advertisements. The advertisements were generated using the five LLMs listed under Section~\ref{paragraph:experiment_dataset} that we will refer to as the set of \emph{old} LLMs. Analogously, we will refer to the prompts used in the WGNA~25 as the \emph{old} prompts. Instead specifying an advertising style as we do in this work, the old prompts define the position of the ad in the response and how it should be phrased, e.g., as a follow-up question or a call-to-action.

For our robustness tests, we created nine new versions of the reference test set by changing 
\Ni
the advertising prompt, 
\Nii 
the LLM used to generate the ad, and 
\Niii
both at the same time.
Each test set variation contains new versions of the 1,904~responses with advertisements. To measure the effect of the prompt and ad-generating LLM on the ad detection effectiveness more accurately, we hold all other variables constant, i.e., the advertised item, its qualities, the advertiser, the response in which the advertisement is inserted, and the user query. The 4,316~responses without advertisements are the same as in the reference test set. The nine test sets result from combining four new prompts with both the old and a new set of LLMs, and from using the old prompts with the new LLMs.

The set of four \emph{new} prompts is based on our taxonomy, each instructing an LLM to apply one of the four advertising styles. To develop the prompts, we took the perspective of advertisers trying to circumvent existing ad-blockers for LLM responses and followed the ad-generation procedure presented in previous works~\cite{zelch:2024, schmidt:2024}. The prompts contain the query posed by the user of a RAG-system as well as the response into which an advertisement should be inserted. For each advertisement, the prompt contains the item, i.e. product or service, that should be advertised, what qualities that item has, as well as the name of the advertiser. Based on the style, the LLM is instructed to use an emotional or rational appeal, and to integrate the ad covertly or overtly into the original response. We improved the prompts iteratively, reducing repeating text patterns and increasing stylistic differences between different categories of our taxonomy. 
In our experiments (Section~\ref{paragraph:results_style_diffs}), we observe significant differences in how effective the classifiers are at detecting advertisements generated by the different prompts. These differences align with our expectations, e.g. covert ads being harder to detect than overt ads, and serve as empirical validation of the successful creation of advertisements in different advertising styles. Figure~\ref{fig:ad-examples} illustrates the process of inserting advertisements into responses and Figure~\ref{fig:prompt-example} shows an example prompt.%
\footnote{All prompts can be found in our repository.}

In addition to the prompts, we varied the ad-generating LLMs to measure their effect on classifier robustness. Based on the set of old LLMs from the WGNA~25 dataset, we selected a set of successors that we will refer to as the \emph{new} LLMs. It consists of \mbox{\texttt{gpt-5-mini}} and \mbox{\texttt{gpt-5-nano}}, the 17B versions of \mbox{\texttt{llama-4-scout}} and \mbox{\texttt{llama-4-maverick}}, and \mbox{\texttt{gpt-oss-120b}}. We accessed the GPT-4 and GPT-5 models via OpenAI and all other LLMs via Groq.%
\footnote{Since the creation of the dataset, Groq has discontinued some of the old LLMs. Following their recommendations, we replaced \mbox{\texttt{deepseek-r1}} and \mbox{\texttt{llama-3-70b}} with \mbox{\texttt{llama-3.3-70b}}, and \mbox{\texttt{qwen-2.5-32b}} with \mbox{\texttt{qwen3-32b}}.}

\paragraph{Measuring Robustness}
\input{figures-and-tables/table-contingency-example.tex}
A robust classifier generalizes to unseen types of advertisements, generated by different LLMs and with different advertising styles. In our experiments, the responses without advertisements are constant between the test sets. Hence, we measure robustness based on the number of ads that a classifier detects. For a given classifier and one of the nine new test sets, we 
\Ni
count the number of true positives and false negatives on the reference test set and the new test set, 
\Nii
add them to a contingency matrix, and 
\Niii
and calculate an odds ratio for ads being detected in responses from the reference test set versus the new test set. 
Table~\ref{tab:contingency_example} shows the contigency table and odds ratio calculation for one classifier and new test set. An odds ratio below 1 indicates that the classifier detected fewer ads on the new than on the reference test set. We consider a classifier robust if the difference in detected ads is not significant. At $\alpha = 0.05$, the difference is significant if the 95\,\% confidence interval of the odds ratio does not include the value 1. We control for a false discovery rate (FDR) of 5\,\% using the Benjamini-Hochberg procedure on $m = 9~\text{Test sets} * 7~\text{Classifiers} = 63~\text{Tests}$.

\subsection{Classifiers}
\label{subsec:experiment_classifiers}
We applied three groups of classifiers to the task of advertisement detection. The first group consists of sentence transformers as applied by \citet{schmidt:2024}. In the second group, we test a new transformer-based approach to ad detection: Classifying tokens to recognize the entities of an advertisement. We expect this higher granularity to be particularly important for advertising implementations like the ones proposed by \citet{dutting:2024} and \citet{soumalias:2024}. Classifiers looking for self-contained segments of text might be less effective if advertisements are spread throughout the response as the result of biased token sampling.

The third and final group consists of conventional, lightweight classifiers in the form of a random forest and a support-vector machine (SVM).
We added this third group for two main reasons: First, we want to analyze how effective ads can be detected without the context used by transformers. Second, ad blockers would ideally run on consumer devices, making efficient inference, both in terms of time and resources, an important feature of ad detectors. As a naive baseline, we also test a dictionary-based approach that assigns probabilities based on manually selected terms from the General Inquirer~\cite{stone:1966}. 

\paragraph{Sentence Classifiers}
\citet{schmidt:2024} used sentence transformers to classify pairs of sentences as containing an advertisement or not. Similar to next sentence prediction, the models were fine-tuned to detect if one sentence followed the other in a response without advertisements.
We reproduced their approach and fine-tuned the same models, \mbox{\texttt{all-MiniLM-L6-v2}} and \mbox{\texttt{all-mpnet-base-v2}}, with the Adam optimizer~\cite{kingma:2014} and binary cross-entropy loss. For MiniLM, we used a batch size of~48 and a learning rate of~1$e$-5. For MPNet, we set the values to~16 and~5$e$-6.
Additionally, we fine-tuned \mbox{\texttt{ModernBERT-embed-base}}%
\footnote{\url{huggingface.co/nomic-ai/modernbert-embed-base}}
on the same task, setting the batch size to~16 and the learning rate to~2$e$-6.
For all three \emph{sentence classifiers}, trained for up to~50 epochs and selected the final weights based on validation F$_1$-score.

\paragraph{Token Classifiers}
Additionally, we fine-tuned two \emph{token classifiers} on the BIO-tags of the WGNA~25. The BIO format is used in named-entity recognition (NER) and other areas of computational linguistics to assign tags to tokens~\cite{ramshaw:1999}. In NER, entities can span multiple tokens. To account for that, the beginning of the entity is marked with a B-tag and all tokens ``in'' the sequence belonging to the entity are marked with I-tags. Other tokens that lie ``outside'' of named entities receive O-tags. Applied to advertisements, the WGNA~25 distinguishes between the item (``B-/I-ITEM''), the advertiser (``B-/I-ADVERTISER''), and the rest of the ad (``B-/I-AD''). All other tokens are tagged as ``O''. 
We fine-tuned \mbox{\texttt{ModernBERT-base}}%
\footnote{\url{https://huggingface.co/answerdotai/ModernBERT-base}}
and \mbox{\texttt{BERT-base-cased}}%
\footnote{\url{https://huggingface.co/google-bert/bert-base-cased}}
on the BIO-tags using the AdamW optimizer~\cite{loshchilov:2019}, a learning rate of~2$e$-5 with linear scheduling, and a batch size of~16. Again, the final weights were chosen based on validation F$_1$-score, this time over 20~epochs. To distinguish sentence classifiers from token classifiers, we give the former a subscript S (e.g. MiniLM$_{\text{S}}$) and the latter a subscript T (MBERT$_{\text{T}}$).

\paragraph{Random Forest}
As a lightweight alternative to transformer-based classifiers, we trained a sentence-level Random Forest classifier~\cite{breiman:2001}, labeling a response as containing an advertisement if at least one sentence is classified as such.
Using scikit-learn~\cite{pedregosa:2011}, we represented each sentence as a binary bag-of-words vector and used a feature selection based on mutual information to retain only the most discriminative terms.
To account for the imbalanced distribution of advertisement and non-advertisement sentences, we used balanced class weights.
We performed a grid search over the minimum document frequency for vocabulary inclusion, the number of selected features, the number of trees, the number of features considered at each split, and the minimum samples per leaf.
We optimized for F$_1$-score on the validation split of the WGNA~25 dataset, tuning the Random Forest's probability threshold to maximize F$_1$-score.

\paragraph{Support Vector Machine}
For comparison, we trained a sentence-level linear SVM, again labeling a response as containing an advertisement if at least one sentence is classified as such.
Each token was represented by a 300-dimensional Word2Vec embedding~\cite{mikolov:2013a}, pretrained on Google News~\cite{mikolov:2013b}, and loaded via Gensim~\cite{rehurek:2010}. Each sentence was represented by the mean of its token embeddings.
We performed a grid search over the regularization parameter $C$, the loss function, and input lowercasing.
The SVM was implemented using scikit-learn~\cite{pedregosa:2011}, calibrated via Platt scaling to obtain probability estimates, and the classification threshold was tuned to maximize F$_1$-score on the validation split.

\paragraph{Dictionary}
Since traditional ad blockers often rely on rule-based systems~\cite{snyder:2020}, and since we observed reoccurring lexical patterns in generated advertisements, we evaluated a dictionary-based approach.
First, we considered words tagged as positive and overstated in the General Inquirer~\cite{stone:1966}, but the resulting classifier performed worse than random predictions.
Second, we selected the 200~words with the highest mutual information toward the advertisement class from the training set. While this was more effective, it still fell short of all other classifiers and is excluded from the following analyses.

%% file: figures-and-tables/table-advertising-dichotomies.tex
\begin{table}[t]
\centering
\small
\setlength{\tabcolsep}{3.5pt}
\renewcommand{\arraystretch}{0.9}
\caption{Confusion matrix of advertising styles found in marketing research: $\blacksquare$ and $\square$ indicate exact and partial overlap.}
\label{table-advertising-dichotomies}
\begin{tabular}{@{}l@{}ll@{\hspace{-15pt}}l@{\hspace{-4pt}}l@{\hspace{-30pt}}l@{\hspace{-30pt}}l@{\hspace{-15pt}}l@{\hspace{-7pt}}l@{\hspace{-6pt}}l@{}}
\toprule
  \multicolumn{2}{@{}l@{}}{\raisebox{1ex}[0em][0em]{\parbox{2cm}{\textbf{\raggedright Marketing style\\dichotomy}}}}
  & \rotatebox{30}{\textbf{Hard-sell}}
  & \rotatebox{30}{\textbf{Soft-sell}}
  & \rotatebox{30}{\textbf{Informational}}
  & \rotatebox{30}{\textbf{Transformational}}
  & \rotatebox{30}{\textbf{Rational}}
  & \rotatebox{30}{\textbf{Emotional}}
  & \rotatebox{30}{\textbf{Overt}}
  & \rotatebox{30}{\textbf{Covert}}
  \\
\midrule 
\multirow{2}{1em}{\rotatebox{90}{\citeyear{mueller:1987}}}                                                                                                     
&  Hard-sell         & $\blacksquare$ &                &   $\square$    &                &   $\square$    &                & $\square$ &                \\
&  Soft-sell         &                & $\blacksquare$ &                &   $\square$    &                &   $\square$    &                & $\square$ \\
\midrule                                                                                                     
\multirow{2}{1em}{\rotatebox{90}{\citeyear{cadet:2017}}}                                                                                                     
&  Informational     &   $\square$    &                & $\blacksquare$ &                &   $\square$    &                &                &                \\
&  Transformational  &                &   $\square$    &                & $\blacksquare$ &                &   $\square$    &                &                \\
\midrule
\multirow{2}{1em}{\rotatebox{90}{\citeyear{kotler:2018}}}                                                                                                     
&  Rational          &   $\square$    &                &   $\square$    &                & $\blacksquare$ &                &                &                \\
&  Emotional         &                &   $\square$    &                &   $\square$    &                & $\blacksquare$ &                &                \\
\midrule
\multirow{2}{1em}{\rotatebox{90}{\citeyear{chan:2019}}}                                                                                                     
&  Overt             & $\square$ &                &                &                &                &                & $\blacksquare$ &                \\
&  Covert            &                & $\square$ &                &                &                &                &                & $\blacksquare$ \\
\bottomrule
\end{tabular}

\end{table}

%% file: figures-and-tables/figure-ad-examples.tex
\begin{figure*}
    \Description{%
    The figure shows a chatbox with a conversation between a user and an LLM. The user asks "Are there specific destinations that usually have last minute deals?". The LLM responds with information about possible destinations.
    Below the chatbox is an item that should be advertised in the response. The item is "FUN Flights" by the advertiser fun.co.uk. The advertisements should emphasize qualities such as "no credit card fees" or "fly from 21 uk airports".
    The remainder of the figure consists of variations of the original LLM response that advertise FUN Flights. 
    The first response is taken from the Webis Generated Native Ads 2025 dataset. Another four responses are grouped in a two-by-two matrix. The first dimension of the matrix is the level of overtness, distinguishing between covert advertisements and overt ones. The second dimension is the type of appeal, i.e., does the advertisement make a rational appeal for the advertised product or does it appeal to emotions.
    }
    \includegraphics[width=0.95\textwidth]{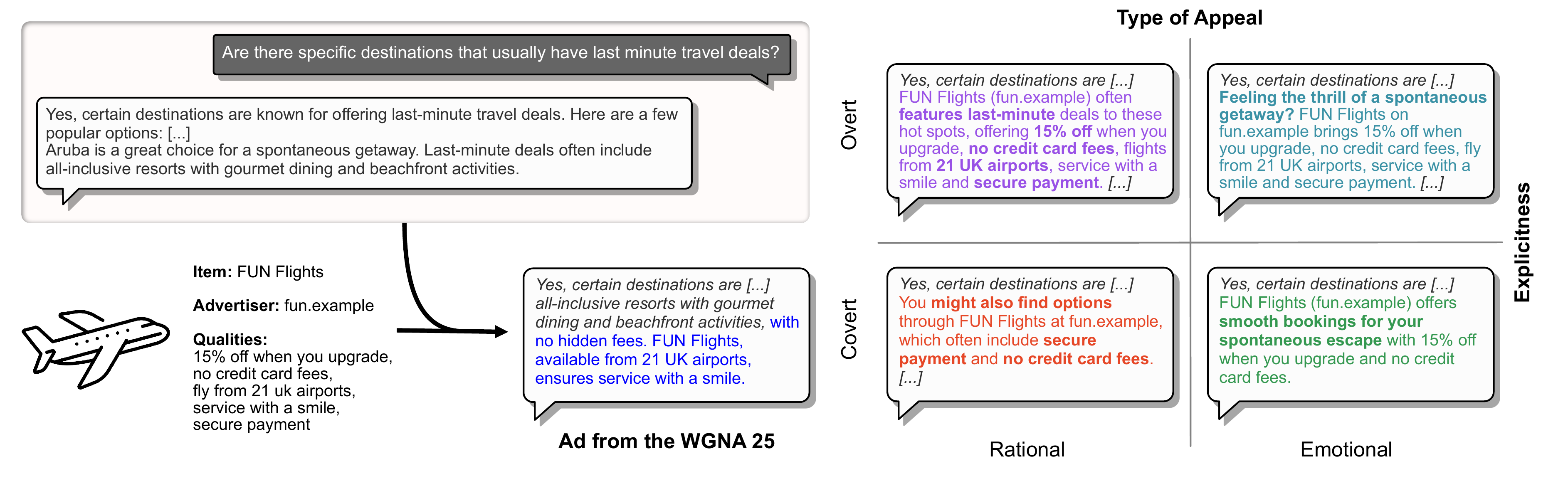}%
    \caption{Example responses for different advertising prompts. The chat window shows a user query for last minute travel and the response generated by a search engine. This response is adapted by inserting an ad for the item ``FUN Flights''. In addition to the response taken from the WGNA~25 test set, the figure shows the variations generated for different advertising styles.}
    \label{fig:ad-examples}
\end{figure*}

%% file: figures-and-tables/table-dataset-overview.tex
\begin{table}[t]
    \caption{%
        Overview of the responses in the Webis Generated Native Ads 2025 dataset
    }
    \label{tab:dataset_overview}
    \begin{minipage}{\columnwidth}
        \centering
        \renewcommand{\tabcolsep}{10pt}
        \begin{tabular}{@{}lrrr|r@{}}
            \toprule
            \bfseries Ad                           &
            \bfseries Train                        &
            \bfseries Val.                         &
            \bfseries Test                         &
            \bfseries Total                                                          \\
            \midrule
            \includegraphics[width=2ex]{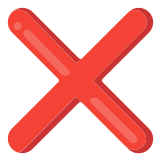}     & 22,416 & 3,999 & 4,316 & 30,731 \\
            \includegraphics[width=2ex]{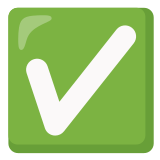} & 10,311 & 1,781 & 1,904 & 13,996 \\
            \midrule
            $\sum$                                 & 32,727 & 5,780 & 6,220 & 44,727 \\
            \bottomrule
        \end{tabular}
        \bigskip
    \end{minipage}
\end{table}

%% file: figures-and-tables/figure-prompt-example.tex
\begin{figure*}[t]
    \Description{%
    The figure shows a prompt instructing an LLM to insert an advertisement into a response.
    The prompt asks the LLM to make rational appeals about the item and produce a covert advertisement.
    }
    \includegraphics[width=0.95\textwidth]{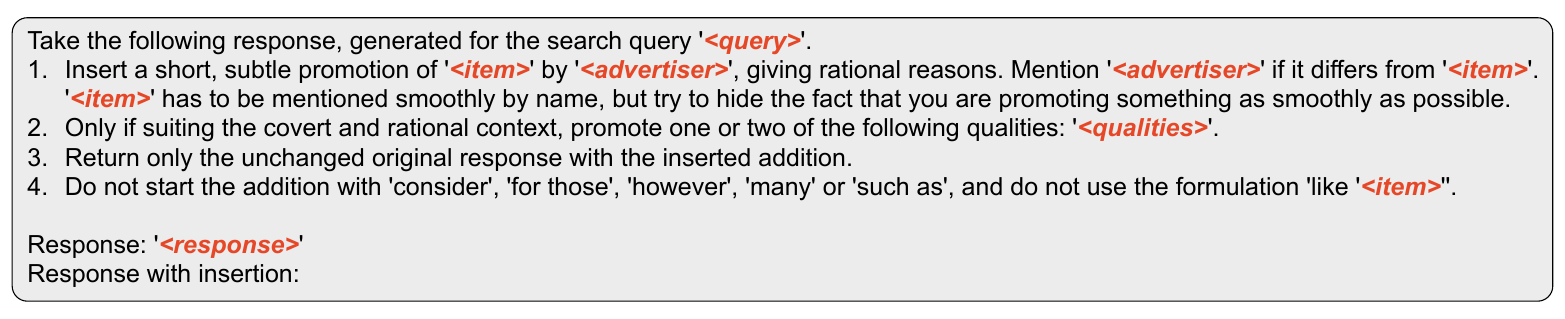}%
    \caption{%
    Prompt to create covert advertisements with rational appeals. The placeholders are filled with the information depicted in Figure~\ref{fig:ad-examples}. 
    }
    \label{fig:prompt-example}
\end{figure*}

%% file: figures-and-tables/table-contingency-example.tex
\begin{table}[t]
    \caption{%
        Contingency table for MiniLM on the test set generated by the covert-emotional prompt with new LLMs
    }
    \label{tab:contingency_example}
    \begin{minipage}{0.47\columnwidth}
        \begin{center}
            \begin{tabular}{l|rr}
                \toprule
                \bfseries Test set   & \bfseries TP & \bfseries FN \\
                \midrule
                New (N)             & 1,355        & 549          \\
                Reference (R)       & 1,785        & 119          \\
                \bottomrule
            \end{tabular}
        \end{center}
        \bigskip
    \end{minipage}
    \hfill
    \begin{minipage}{0.5\columnwidth}
        \begin{math}
            \text{OR} = \frac{\text{TP}_\text{N} / \text{FN}_\text{N}}{\text{TP}_\text{R} / \text{FN}_\text{R}} = \frac{1355 / 549}{1785 / 119}
        \end{math}
        
        \vspace{2ex}
        \begin{math}
            \text{OR} = 0.165, \text{CI}_{0.95} = [0.13, 0.20]
        \end{math}
        \bigskip
    \end{minipage}
\end{table}

%% file: sigir26-ads-in-rag-part4.tex
\section{Evaluation}
\label{sec:results}

The following section describes our evaluation of the classifiers' effectiveness on the test data and their robustness to changes in the LLMs and prompts used for generating the advertisements. We find significant differences between advertising styles, and that effectiveness is largely correlated with the parameter count of the detector. We further test the token-based classifiers for their ability to detect named entities relating to the generated advertisements.

\subsection{Classifier Effectiveness}
\input{figures-and-tables/table-transformer-effectiveness.tex}
\input{figures-and-tables/table-context-free-effectiveness.tex}

As a first step, we evaluated each classifier on the reference test set generated with the ``old'' LLMs and prompts. The results for all classifiers are given in the first row of Tables~\ref{tab:transformer_effectiveness} and~\ref{tab:context_free_effectiveness}. Aside from the SVM, all classifiers achieve F\textsubscript{1}-scores of 0.9 and higher.
We observe that
\Ni 
the token classifiers are more effective than their counterparts that classify pairs of sentences, and 
\Nii
the classification effectiveness increases with the number of parameters.

Tables~\ref{tab:transformer_effectiveness} and \ref{tab:context_free_effectiveness} also summarize the effectiveness scores on all nine new test sets, created by varying the LLM, the prompt, or both.
As the negative class (responses without ads) contains the same examples in all test sets, any deviation in a classifier's precision stems entirely from a decrease in true positives.
The recall scores on both the reference and the new test sets show the same trends: Models with more parameters are more effective than smaller models, and token classifiers are more effective than sentence classifiers.
Surprisingly, we observe that all transformer-based classifiers are actually \emph{better} at identifying ads generated by the overt-emotional prompt than those from the reference set, which they were trained on. The only exception is the sentence classifier MiniLM (MiniLM\textsubscript{S}), which is slightly less effective on overt-emotional advertisements generated by new LLMs.

\subsection{Classifier Robustness}
\input{figures-and-tables/figure-odds-ratio.tex}
To test if a classifier is robust to changes in the LLM or prompt, we calculated the odds ratio (odds being true positives against false negatives) in a 95\,\% confidence interval between ads detected in responses from the reference test set versus each of the new test sets (illustrated in Table~\ref{tab:contingency_example}). The difference in the number of detected ads is significant at $\alpha=0.05$ when the confidence interval does not contain~1.
The confidence intervals are depicted in Figure~\ref{fig:odds-ratio}. All significant differences (controlled for a false discovery rate of 5\,\%) are marked with an asterisk in Tables~\ref{tab:transformer_effectiveness} and \ref{tab:context_free_effectiveness}.
The results show that most classifiers detect significantly fewer ads in most of the new test sets, and are thus not robust to the respective change in prompt and\,/\,or LLM. The one exception is the token-classifier based on ModernBERT (MBERT\textsubscript{T}), that is robust to all changes except the covert advertisements generated by new LLMs. Both the random forest and the SVM are not robust to any of the changes. Similar to the transformers, however, they achieve the highest effectiveness scores on the overt-emotional advertisements generated by the set of old LLMs. The SVM is overall less effective than the random forest, but generalizes better to the new test sets.

\paragraph{Differences Between Advertising Styles}
\label{paragraph:results_style_diffs}
The odds ratios in Figure~\ref{fig:odds-ratio} show that the covert advertising style is more difficult to detect than the overt style. 
Most classifiers are significantly less effective at detecting these advertisements. While MBERT\textsubscript{T} is good at detecting the covert advertising style generated by LLMs also used in the reference test set, the model is significantly less effective on the same advertising style generated by new LLMs.
In contrast, the overt-emotional advertising style yields very different results: MBERT\textsubscript{T} is able to detect \emph{all} advertisements generated in this style by the old LLMs and almost all generated by new LLMs. The BERT token classifier (BERT\textsubscript{T}), the sentence classifiers based on ModernBERT (MBERT\textsubscript{S}), and MPNET (MPNET\textsubscript{S}) are less effective on the reference test set, but detect significantly more ads than in the reference test set.
For the overt-rational prompt, the results are less consistent. All sentence classifiers detect significantly fewer ads in this advertising style, independent of the LLM. The token classifiers all detect fewer of the ads generated by new LLMs; for BERT\textsubscript{T} the reduction is significant, for MBERT\textsubscript{T} it is not.

To quantify the differences between advertising styles, we performed the same odds ratio calculation as for the robustness analysis. We hold everything constant except for the dimension of interest and compare the number of detected ads between related pairs. For the comparison covert vs. overt, 
this means comparing the results for the same classifier, the same set of LLMs, and the same type of appeal (emotional or rational). 
This results in 7~Classifiers\,\texttimes{}\,2~LLM Sets\,\texttimes{}\,2~Advertising Styles\,=\,28~Comparisons, again controlled for an FDR of 5\,\%.
In 26 out of those 28 comparisons, the advertisements with an overt style are significantly easier to detect than their covert counterparts. In the other two comparisons, the overt advertisements are also easier to detect but not significantly so. Comparing the emotional against the rational appeal, we find the emotional advertisements to be significantly easier to detect in 20 of 28~cases.

\paragraph{Differences Between LLMs}
Similar to the advertising styles, we tested if the classifiers are robust to using a different set of LLMs to generate the advertisements. Looking at Tables~\ref{tab:transformer_effectiveness} and~\ref{tab:context_free_effectiveness}, we see that all sentence transformers, the random forest, and the SVM detect significantly fewer ads when combining the old set of prompts with the new set of LLMs. The two token classifiers, however, are robust against this change.
Using the same approach as for the advertising styles above, we find that in~17 of~28 comparisons, the advertisements generated with the set of old LLMs are significantly easier to detect than their counterparts generated with new LLMs.

\paragraph{Overlap in False Negatives}
\input{figures-and-tables/figure-false-negative-overlap.tex}
Additionally, we compared the advertisements that the classifiers did not detect. For each of the nine new test sets and the reference test set, we formed pairs of classifiers and calculated the Jaccard index of their false negatives. Figure~\ref{fig:false_negative_overlap} depicts the average Jaccard index for each pair across all test sets.
The two sentence classifiers MPNET\textsubscript{S} and MiniLM\textsubscript{S} have the highest overlap in false negatives with an average Jaccard index of~0.452. We observe that most classifiers have the highest overlap with other classifiers in the same category: The random forest has the highest overlap with the SVM (and vice versa), the sentence classifier MBERT\textsubscript{S} has the highest overlap with MPNET\textsubscript{S}, and the token classifier MBERT\textsubscript{T} has the highest overlap with BERT\textsubscript{T}. The only exception is the token classifier BERT\textsubscript{T}, which has its highest overlap with the sentence classifier MBERT\textsubscript{S}.

\subsection{Effectiveness of Entity Recognition}
\input{figures-and-tables/table-entity-effectiveness}
\input{figures-and-tables/figure-odds-ratio-entities.tex}
In addition to the binary classification task of labeling a response as containing an ad or not, we also evaluated the token classifiers for their ability to detect the following three entities: \emph{item}, \emph{advertiser}, and \emph{other advertising text} (such as item qualities). These entities are marked with BIO-tags at token-level. The tags signal if a token marks the \emph{beginning} of an entity (e.g., ``B-ITEM''), occurs \emph{inside} a sequence of tokens belonging to the entity (e.g., ``I-ADVERTISER''), or \emph{outside} of it (``O''). We calculated the effectiveness of the two token classifiers using the \texttt{seqeval} metric of the \texttt{evaluate} Python package.%
\footnote{\url{https://github.com/huggingface/evaluate}}
The values for precision and recall in Table~\ref{tab:entity_effectiveness} are based on 
\Ni 
all ground-truth entities in the respective test set and 
\Nii 
all entity predictions by the respective classifier. 
To detect an entity, the classifier needs to assign the correct BIO-tags to \emph{all} tokens belonging to that entity. If fewer or more tokens are tagged, the entity is counted as a false negative. It is important to mention that, in contrast to the response classification, the precision scores vary between test sets. This is because, unlike before, the number of false positives can also increase for the newly generated responses.

To perform similar robustness tests as for the response classification, we counted the number of responses for which a classifier detected \emph{all} entities. We performed this aggregation, because the odds ratio test requires each trial to be independent of previous trials. However, this requirement may not hold if we treat the all entity detections as individual trials, since the entities themselves are not necessarily independent. If one entity is detected in a response, the probability of detecting other entities in the same response might increase. This macro aggregation is therefore stricter than the micro effectiveness scores in Table~\ref{tab:entity_effectiveness}, but necessary to perform the test. We did not add asterisks to Table~\ref{tab:entity_effectiveness}, as the basis for the recall scores is hence different from the other odds ratio calculations.
We treated a response as a true positive if the classifier detected \emph{all} its entities, and as a false negative otherwise. With these counts, the odds ratios were calculated in the same way by comparing the odds from the new test sets against those from the reference test set. The resulting odds ratios and their corresponding 95\,\% confidence intervals are depicted in Figure~\ref{fig:odds-ratio-entities}. 

We observe the same tendencies as for the ad detection: 
First, the token classifiers correctly detect more entities when the advertisements were generated in an overt instead of a covert advertising style.
Second, the entities in advertisements with an emotional appeal are easier to detect than in advertisements with a rational appeal.
The detection odds ratios, however, are not as clear. The only prompt with consistent results is the covert-rational prompt, for which both classifiers detect significantly fewer entities, regardless of the generating LLM. ModernBERT is not robust to the covert-emotional prompt, while BERT is robust when the ads are generated by the set of old LLMs. 

The overt-rational prompts lead to no significant changes in the entity detection. BERT is significantly \emph{more} effective on the overt-emotional advertisements generated by the old LLMs and robust to those generated by the new LLMs. ModernBERT, however, is not robust to the new LLMs generating overt-emotional advertisements. Overall, ModernBERT is less robust at detecting advertising entities than on the task of detecting responses with advertisements.

%% file: figures-and-tables/table-transformer-effectiveness.tex
\begin{table*}[t]
    \caption{%
        Effectiveness of the transformer-based classifiers. The colors indicate if the classifier detects more, fewer or the same number of ads as on the reference test set (old LLMs \& prompts). Asterisks (*) highlight significant differences. Classifiers with a subscript S classify pairs of sentences (MiniLM$_\text{S}$), those with a T classify tokens (MBERT$_\text{T}$).
    }
    \label{tab:transformer_effectiveness}
    \centering
    \small
    \begin{tabular}{@{}llrrrrrrrrr|rrrrrr@{}}
    \toprule
    & & 
        \multicolumn{3}{c}{\bfseries MiniLM$_\text{S}$} & 
        \multicolumn{3}{c}{\bfseries MPNET$_\text{S}$}  & 
        \multicolumn{3}{c}{\bfseries MBERT$_\text{S}$}  & 
        \multicolumn{3}{c}{\bfseries BERT$_\text{T}$}   & 
        \multicolumn{3}{c}{\bfseries MBERT$_\text{T}$}  \\
    \cmidrule(lr){3-5} \cmidrule(lr){6-8} \cmidrule(lr){9-11} \cmidrule(lr){12-14} \cmidrule(lr){15-17}
    \bfseries LLMs & \bfseries Prompt & 
        \multicolumn{1}{c}{Prec.} & \multicolumn{1}{c}{Rec.} & \multicolumn{1}{c}{F$_1$} & 
        \multicolumn{1}{c}{Prec.} & \multicolumn{1}{c}{Rec.} & \multicolumn{1}{c}{F$_1$} & 
        \multicolumn{1}{c}{Prec.} & \multicolumn{1}{c}{Rec.} & \multicolumn{1}{c}{F$_1$} & 
        \multicolumn{1}{c}{Prec.} & \multicolumn{1}{c}{Rec.} & \multicolumn{1}{c}{F$_1$} & 
        \multicolumn{1}{c}{Prec.} & \multicolumn{1}{c}{Rec.} & \multicolumn{1}{c}{F$_1$} \\
    \midrule
    \multirow[c]{5}{*}{Old} & \textit{Old Prompts} & \textit{0.976} & \textit{0.938} & \textit{0.957} & \textit{0.992} & \textit{0.926} & \textit{0.958} & \textit{0.964} & \textit{0.977} & \textit{0.971} & \textit{0.987} & \textit{0.983} & \textit{0.985} & \bfseries \textit{0.995} & \bfseries \textit{0.998} & \textit{0.997} \\
    \cmidrule(lr){2-17}
                            & Overt-Emotional & 0.977 & \highercol 0.959\rlap{*} & 0.968 & 0.992 & \highercol 0.962\rlap{*} & 0.977 & 0.965 & \highercol 0.988\rlap{*} & 0.977 & 0.987 & \highercol 0.992\rlap{*} & 0.989 & \bfseries 0.995 & \highercol \bfseries 1.000 & 0.997 \\
                            & Overt-Rational & 0.975 & \lowercol 0.868\rlap{*} & 0.918 & 0.991 & \lowercol 0.847\rlap{*} & 0.914 & 0.963 & \lowercol 0.946\rlap{*} & 0.955 & 0.987 & \lowercol 0.981 & 0.984 & \bfseries 0.995 & \noeffect \bfseries 0.998 & 0.997 \\
                            & Covert-Emotional & 0.973 & \lowercol 0.799\rlap{*} & 0.877 & 0.991 & \lowercol 0.780\rlap{*} & 0.873 & 0.962 & \lowercol 0.925\rlap{*} & 0.944 & 0.986 & \lowercol 0.951\rlap{*} & 0.968 & \bfseries 0.995 & \lowercol \bfseries 0.996 & 0.995 \\
                            & Covert-Rational & 0.969 & \lowercol 0.709\rlap{*} & 0.819 & 0.989 & \lowercol 0.651\rlap{*} & 0.785 & 0.960 & \lowercol 0.855\rlap{*} & 0.904 & 0.986 & \lowercol 0.943\rlap{*} & 0.964 & \bfseries 0.995 & \lowercol \bfseries 0.995 & 0.995 \\
    \cmidrule(lr){1-17}
    \multirow[c]{5}{*}{New} & Old Prompts & 0.975 & \lowercol 0.871\rlap{*} & 0.920 & 0.992 & \lowercol 0.897\rlap{*} & 0.942 & 0.964 & \lowercol 0.962\rlap{*} & 0.963 & 0.987 & \lowercol 0.979 & 0.983 & \bfseries 0.995 &  \lowercol \bfseries 0.998 & 0.996 \\
    \cmidrule(lr){2-17}
                            & Overt-Emotional & 0.976 & \lowercol 0.932 & 0.954 & 0.992 & \highercol 0.949\rlap{*} & 0.970 & 0.965 & \highercol 0.992\rlap{*} & 0.978 & 0.987 & \highercol 0.991\rlap{*} & 0.989 & \bfseries 0.995 & \highercol \bfseries 0.999 & 0.997 \\
                            & Overt-Rational & 0.973 & \lowercol 0.805\rlap{*} & 0.881 & 0.991 & \lowercol 0.809\rlap{*} & 0.891 & 0.963 & \lowercol 0.942\rlap{*} & 0.952 & 0.987 & \lowercol 0.973\rlap{*} & 0.980 & \bfseries 0.995 &  \lowercol \bfseries 0.997 & 0.996 \\
                            & Covert-Emotional & 0.969 & \lowercol 0.712\rlap{*} & 0.821 & 0.990 & \lowercol 0.743\rlap{*} & 0.849 & 0.961 & \lowercol 0.902\rlap{*} & 0.931 & 0.986 & \lowercol 0.926\rlap{*} & 0.955 & \bfseries 0.995 & \lowercol \bfseries 0.988\rlap{*} & 0.992 \\
                            & Covert-Rational & 0.965 & \lowercol 0.627\rlap{*} & 0.760 & 0.988 & \lowercol 0.592\rlap{*} & 0.741 & 0.957 & \lowercol 0.808\rlap{*} & 0.876 & 0.986 & \lowercol 0.905\rlap{*} & 0.944 & \bfseries 0.995 & \lowercol \bfseries 0.989\rlap{*} & 0.992 \\
    \bottomrule
    \end{tabular}
\end{table*}

%% file: figures-and-tables/table-context-free-effectiveness.tex
\begin{table}[t]
    \caption{%
        Effectiveness of the context-free classifiers. Significant differences are highlighted with an asterisk (*).
    }
    \label{tab:context_free_effectiveness}
    \centering
    \small
    \setlength{\tabcolsep}{5pt}
    \begin{tabular}{@{}llrrrrrr@{}}
    \toprule
    & & 
        \multicolumn{3}{c}{\bfseries SVM} & 
        \multicolumn{3}{c}{\bfseries RForest} \\
    \cmidrule(lr){3-5} \cmidrule(lr){6-8}

    \bfseries LLMs & \bfseries Prompt &
        \multicolumn{1}{c}{Prec.} & \multicolumn{1}{c}{Rec.} & \multicolumn{1}{c}{F$_1$} & 
        \multicolumn{1}{c}{Prec.} & \multicolumn{1}{c}{Rec.} & \multicolumn{1}{c}{F$_1$} \\
    \midrule
    \multirow[c]{5}{*}{Old} & \textit{Old Prompts}  & \textit{0.470} & \textit{0.812} & \textit{0.595} & \textit{0.883} & \textit{0.914} & \textit{0.898} \\
    \cmidrule(lr){2-8}
                            & Overt-Emo.            & 0.453 & \lowercol 0.758\rlap{*} & 0.567 & 0.870 & \lowercol 0.811\rlap{*} & 0.840 \\
                            & Overt-Rat.            & 0.434 & \lowercol 0.703\rlap{*} & 0.537 & 0.861 & \lowercol 0.751\rlap{*} & 0.802 \\
                            & Covert-Emo.           & 0.374 & \lowercol 0.547\rlap{*} & 0.444 & 0.798 & \lowercol 0.478\rlap{*} & 0.598 \\
                            & Covert-Rat.           & 0.370 & \lowercol 0.539\rlap{*} & 0.439 & 0.792 & \lowercol 0.461\rlap{*} & 0.582 \\
    \cmidrule(lr){1-8}
    \multirow[c]{5}{*}{New} & Old Prompts           & 0.434 & \lowercol 0.703\rlap{*} & 0.536 & 0.856 & \lowercol 0.721\rlap{*} & 0.782 \\
    \cmidrule(lr){2-8}
                            & Overt-Emo.            & 0.441 & \lowercol 0.724\rlap{*} & 0.548 & 0.854 & \lowercol 0.711\rlap{*} & 0.776 \\
                            & Overt-Rat.            & 0.406 & \lowercol 0.627\rlap{*} & 0.493 & 0.848 & \lowercol 0.674\rlap{*} & 0.751 \\
                            & Covert-Emo.           & 0.376 & \lowercol 0.552\rlap{*} & 0.447 & 0.780 & \lowercol 0.431\rlap{*} & 0.555 \\
                            & Covert-Rat.           & 0.357 & \lowercol 0.508\rlap{*} & 0.419 & 0.780 & \lowercol 0.430\rlap{*} & 0.554 \\
    \bottomrule
    \end{tabular}
\end{table}

%% file: figures-and-tables/figure-odds-ratio.tex
\begin{figure*}
    \Description{%
    The figure shows the odds ratios for detecting advertisements on new test sets vs. the reference test set. The odds ratios are divided across four graphs, one for each advertising style. From left to right, the plots show the odds ratios for the overt-emotional, the overt-rational, the covert-emotional, and the covert-rational advertising style. The odds ratios are shown on the x-axis together with their 95\,\% confidence interval. The y-axis lists the different classifiers used in the experiments.
    The odds ratios for the overt-emotional advertising style are greater than 1 for a lot of classifiers. For both of the covert advertising styles, the odds ratios are smaller than 1. The overt-rational advertising style shows a mixed result: the token classifiers have odds ratios lower than 1 with intervals overlapping 1, while other classifiers have odds ratios that are significantly different from 1.   
    }
    \includegraphics[width=0.95\textwidth]{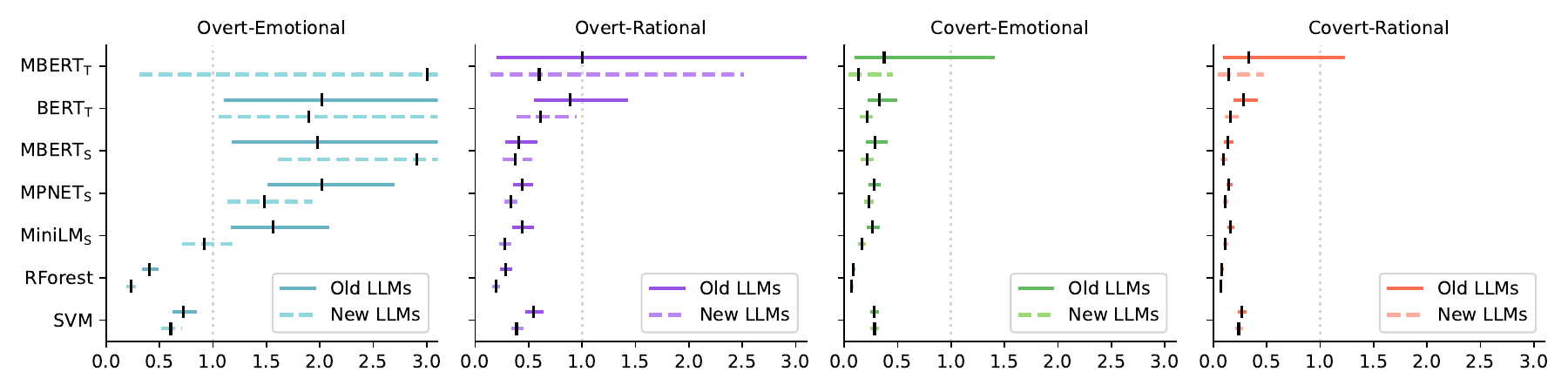}%
    \caption{Ad detection odds ratios (95\,\% CI). For each classifier and new test set, we compared the odds of detecting an ad in the new test set to the odds in the reference test set (see also Table~\ref{tab:contingency_example}). The black vertical ticks show the odds ratio and the colored horizontal lines the corresponding confidence interval. The x-Axis is cut at 3.0 for improved clarity.}
    \label{fig:odds-ratio}
\end{figure*}

%% file: figures-and-tables/figure-false-negative-overlap.tex
\begin{figure}
    \Description{%
    The figure shows a heatmap for the average overlap in false negatives between different classifiers, measured by a Jaccard index. The highest overlap is between MPNET and MiniLM with 0.452. Generally, the classifiers in the same category (token classifiers, sentence classifiers, and context-free classifiers) have a higher overlap with each other than with classifiers from other categories. 
    }
    \includegraphics[width=0.47\textwidth]{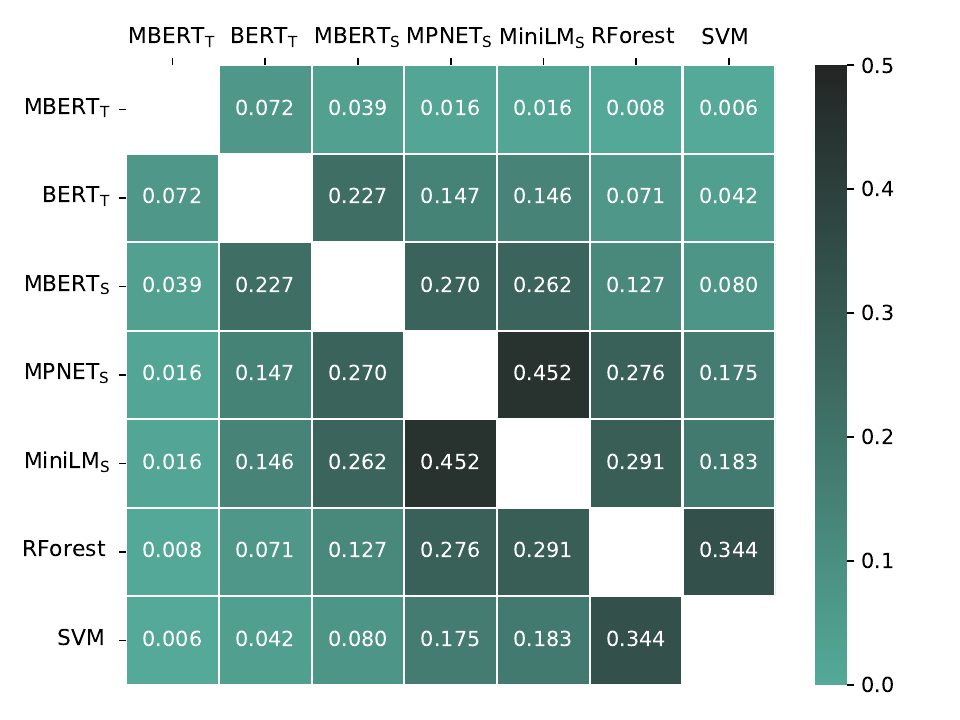}%
    \caption{Average overlap in false negatives. The heatmap shows the mean Jaccard index over all test sets. 
    A score of 1 indicates that two classifiers always miss the same ads.}
    \label{fig:false_negative_overlap}
\end{figure}

%% file: figures-and-tables/table-entity-effectiveness.tex
\begin{table}[t]
    \caption{%
        Effectiveness on entity recognition. Scores are calculated based on all labeled and detected entities. We do not report significance, as the detection of entities in the same response cannot be assumed to be independent events. 
    }
    \label{tab:entity_effectiveness}
    \centering
    \small
    \setlength{\tabcolsep}{5pt}
    \begin{tabular}{@{}llrrrrrr@{}}
    \toprule
    & & 
        \multicolumn{3}{c}{\bfseries BERT$_\text{T}$} & 
        \multicolumn{3}{c}{\bfseries MBERT$_\text{T}$} \\
    \cmidrule(lr){3-5} \cmidrule(lr){6-8}

    \bfseries LLMs & \bfseries Prompt &
        \multicolumn{1}{c}{Prec.} & \multicolumn{1}{c}{Rec.} & \multicolumn{1}{c}{F$_1$} & 
        \multicolumn{1}{c}{Prec.} & \multicolumn{1}{c}{Rec.} & \multicolumn{1}{c}{F$_1$} \\
    \midrule
    \multirow[c]{5}{*}{Old} & \textit{Old Prompts}   & \textit{0.698} & \textit{0.757} & \textit{0.726} & \bfseries \textit{0.778} & \bfseries \textit{0.818} & \textit{0.798} \\
    \cmidrule(lr){2-8}
                            & Overt-Emo.    & 0.756 & \highercol 0.772 & 0.764 & \bfseries 0.813 & \bfseries \highercol 0.822 & 0.817 \\
                            & Overt-Rat.    & 0.704 & \lowercol  0.725 & 0.714 & \bfseries 0.784 & \bfseries \lowercol  0.801 & 0.793 \\
                            & Covert-Emot.  & 0.634 & \lowercol  0.667 & 0.650 & \bfseries 0.696 & \bfseries \lowercol  0.726 & 0.711 \\
                            & Covert-Rat.   & 0.609 & \lowercol  0.632 & 0.620 & \bfseries 0.674 & \bfseries \lowercol  0.714 & 0.694 \\
    \cmidrule(lr){1-8}
    \multirow[c]{5}{*}{New} & Old Prompts   & 0.659 & \lowercol  0.696 & 0.677 & \bfseries 0.739 & \bfseries \lowercol  0.760 & 0.750 \\
    \cmidrule(lr){2-8}
                            & Overt-Emot.   & 0.747 & \lowercol  0.743 & 0.745 & \bfseries 0.786 & \bfseries \lowercol  0.789 & 0.787 \\
                            & Overt-Rat.    & 0.699 & \lowercol  0.697 & 0.698 & \bfseries 0.758 & \bfseries \lowercol  0.773 & 0.766 \\
                            & Covert-Emot.  & 0.657 & \lowercol  0.641 & 0.649 & \bfseries 0.707 & \bfseries \lowercol  0.707 & 0.707 \\
                            & Covert-Rat.   & 0.605 & \lowercol  0.585 & 0.595 & \bfseries 0.688 & \bfseries \lowercol  0.694 & 0.691 \\
    \bottomrule
    \end{tabular}
\end{table}

%% file: figures-and-tables/figure-odds-ratio-entities.tex
\begin{figure}
    \Description{%
    The figure shows the odds ratios for detecting the entities in advertisements on new test sets vs. the reference test set. The odds ratios are shown for the token classifiers based on ModernBERT and BERT. The y-axis lists the combinations of classifiers and test set: At the top is ModernBERT on the test set created with the overt-emotional advertising style. Next is BERT for the same test set. This is continued for the overt-rational, the covert-emotional, and the covert-rational advertising style. 
    The odds ratios are shown on the x-axis together with their 95\,\% confidence interval. 
    Only two odds ratios are larger than one, namely those for both classifiers on the test set created by old LLMs with the overt-emotional advertising style. Among those, only the one for BERT is significantly larger than 1.
    The remaining odds ratios are smaller than one, and all except for one are significantly smaller. The one exception is BERT on the test set created by new LLMs with the overt-emotional advertising style.
    }
    \centering
    \begin{minipage}{0.46\textwidth}
        \includegraphics[width=\linewidth]{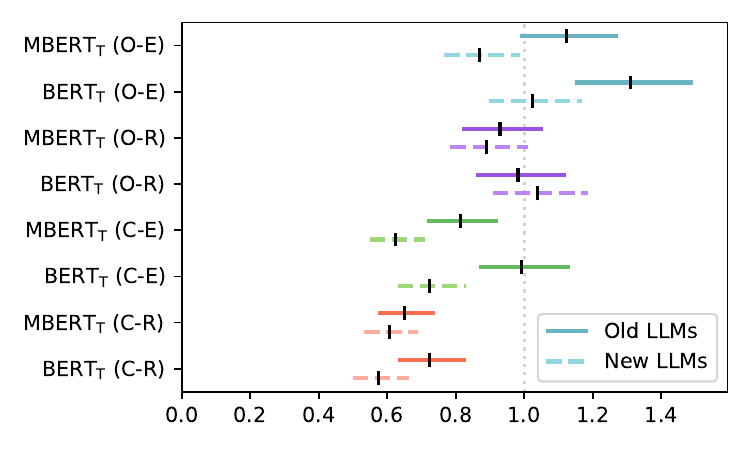}%
    \end{minipage}
    \caption{Entity recognition odds ratios (95\,\% CI). For each classifier and new test set, we compared the odds of detecting \emph{all} entities in a response against the odds in the reference test set (Differs from the entity-level effectiveness in Table~\ref{tab:entity_effectiveness}).}
    \label{fig:odds-ratio-entities}
\end{figure}

%% file: sigir26-ads-in-rag-part5.tex
\section{Discussion and Limitations}

Our results indicate that many classifiers are not robust to changes in the advertising style or the ad-generating LLM. The only exception is the token classifier based on ModernBERT that retains a lot of its effectiveness across all nine new test sets. The lowest recall value, 0.988, occurs for the advertisements generated by the set of new LLMs with a covert-emotional advertising style. In absolute terms, ModernBERT misses 22~responses with advertisements in this test set, compared to 3~responses in the reference test set.

In line with their definition, covert advertisements are harder to detect than their overt counterparts. We also observe that emotional appeals seem to be easier to detect than rational ones. One possible explanation is that the ad-generating LLM needs to use additional vocabulary to create an emotional narrative, making it easier for classifiers to detect.
We also find that the set of new LLMs are better at generating advertisements that evade detection by our classifiers.  This could be either due to a better adherence to the covert advertising style or a general shift in their generated advertisements that removes patterns that the classifiers look for. In either case, it underlines the need for robust ad detectors. 

We find classifiers operating at token-level to be both more effective and more robust than classifiers operating on larger units of text. A possible explanation is that the context of the full response allows these classifiers to evaluate each token in interaction with all other tokens, thereby detecting patterns like a positive adjective referring to a product name.
At the same time, these token classifiers are not as effective or robust on the task of detecting advertising entities. This task, however, plays an important role in the development of ad blockers for LLMs: Only with an exact position of the advertising text can ad blockers be precise and not remove too much of a generated response.
Although the context-free classifiers that we tested are not as reliable as the transformer-based models, they might still find legitimate use in the creation of production-ready ad blockers. The random forest classifier is effective at detecting similar advertisements to the ones it was trained on and end-user devices like smartphones cannot be guaranteed to fulfill the hardware requirements of the transformer models.

Overall, our results are limited to the data, LLMs, and classifiers we used in our experiments. With regard to the data, this includes the specific advertisements of the WGNA~25 dataset with their structure of item, qualities, and advertiser. Furthermore, the advertisements were inserted \emph{after} a response was already generated. In future work, we want to implement advertising mechanisms that generate the response from an advertiser-biased distribution and study the robustness of classifiers to this setup.

%% file: sigir26-ads-in-rag-sum.tex
\section{Conclusion}

Advertisements will become a part of RAG responses. In preparation for this scenario, existing research has explored the detection of ads in the response texts. 
We contribute to this emerging field of research by
\Ni
proposing a taxonomy of advertising styles for LLMs
\Nii
simulating that advertisers may evade trained ad detectors by changing their advertising style, and 
\Niii
studying the robustness of various ad detectors under these changes.

We find classifiers operating on individual tokens to be both more effective and more robust than classifiers proposed by previous research. Especially ModernBERT is consistently effective at detecting responses with advertisements. Conventional classifiers like a random forest and SVM are not robust to changes in the advertising style, suggesting that lexical patterns alone are insufficient for a reliable detection, and that context information is important.
Finally, our experiments reveal room for improvement in the task of precisely locating an ad in a response. Future work in this area will be a valuable step towards ad blockers for LLMs and thus contribute to higher informational quality in RAG responses.

\begin{acks}
This publication has received funding from the European Union's Horizon Europe research and innovation programme under grant agreement No 101070014 (OpenWebSearch.EU, \url{https://doi.org/10.3030/101070014}). 
\end{acks}